\documentclass[aps,10pt]{revtex4}
\def\theequation{\arabic{section}.\arabic{equation}}
\usepackage{psfrag}
\usepackage{subfigure}
\usepackage{color}
\usepackage{mathrsfs}
\usepackage{graphicx}
\usepackage[colorlinks=true,linkcolor=blue,citecolor=magenta,urlcolor=blue]{hyperref}
\usepackage{amssymb, bm}
\usepackage{amsmath, amsthm}
\usepackage{epstopdf}
\usepackage{hyperref}
\usepackage{enumerate}
\usepackage{longtable}
\usepackage{amsmath}

\usepackage{amsmath}  
\usepackage{amsfonts} 
\usepackage{graphicx}

\newcommand{\be}{\begin{equation}}
\newcommand{\ee}{\end{equation}}

\newcommand{\ms}{\medskip}

\begin{document}
\def\theequation{\arabic{section}.\arabic{equation}}

\title{Singular parametric oscillators from the one-parameter Darboux transformation of the classical harmonic oscillator}
\author{H.C. Rosu}
\email{hcr@ipicyt.edu.mx; ORCID: 0000-0001-5909-1945} 
\affiliation{Instituto Potosino de Investigaci\'on Cient\'{\i}fica y Tecnol\'ogica, Camino a la Presa San Jos\'e 2055,
Col. Lomas 4a Secci\'on, San Luis Potos\'{\i}, 78216 S.L.P., Mexico}

\author{J. de la Cruz}
\email{josue.delacruz@ipicyt.edu.mx; ORCID: 0000-0001-5943-5752}
\affiliation{Instituto Potosino de Investigaci\'on Cient\'{\i}fica y Tecnol\'ogica, Camino a la Presa San Jos\'e 2055,
Col. Lomas 4a Secci\'on, San Luis Potos\'{\i}, 78216 S.L.P., Mexico}

\centerline{Ann. Phys. 470 (2024) 169830}
\centerline{doi:10.1016/j.aop.2024.169830}

\date{2403.12084v2}

\bigskip
\bigskip
\begin{abstract}
The singular parametric oscillators obtained from the
one-parameter Darboux deformation/transformation effected upon the classical harmonic oscillator are introduced and discussed in some detail
using $\sin(\omega_0 t)$ and $\cos(\omega_0 t)$ as seed solutions.
The corresponding Ermakov-Lewis integrability problem of these parametric oscillators is also studied. It is shown that their Ermakov-Lewis
invariants do not depend on the deformation parameter and are singularity-free.

\medskip


\end{abstract}


\maketitle

\section{Introduction}\label{sec:1}
\setcounter{equation}{0}

One-parameter Darboux transformations of homogeneous second-order linear differential equations have been an interesting topic in the literature since four decades now. Different from the common Darboux transformations which are based on a particular Riccati solution, they are based on the general Riccati solution and therefore generate whole families of Darboux covariant equations. In a quantum mechanical context, this covariance has been used to generate many exactly solvable Schr\"odinger problems that are isospectrally related to a given exactly solvable quantum eigenvalue problem with the parameter related to the changed character of the boundary conditions. This kind of parametric Darboux transformations has been applied in many areas of physics \cite{pM,pF,psp,cks,rosu96-1,rosu96-2,rosu97,r2000,RMC,KKYK} including the free damped harmonic oscillator in classical mechanics \cite{rr98,re01}, but, to the best of our knowledge, the pure harmonic oscillator has not been studied in this approach. Mathematically, the parametric Darboux transformations turn these simple autonomous equations into nonautonomous equations pertaining to the vast area of time-dependent oscillators of ever lasting research interest \cite{td1, td2,td3}. On the other hand, at the formalism level, supersymmetric classical Hamiltonians and Lagrangians have been studied in the literature employing additional Grassmannian variables \cite{bagchibook,bissoni86,ioffe06}.

\medskip

In this paper, our aim is to provide a detailed study of the parametric oscillators obtained by applying the one-parameter Darboux transformation to the classical harmonic oscillator,
\be\label{odeG}
F''+\omega_0^2F=0 ~,
\ee
where $\omega_0$ is the natural frequency of the oscillator as a trigonometric counterpart of the hyperbolic `oscillator'
\be\label{odeF}
y_i''-\tilde{\varphi}^2y_i=0 ~,
\ee
as discussed in the one-parameter supersymmetric framework in the recent work \cite{rm23}. In (\ref{odeF}),
$\tilde{\varphi}\approx 0.4812$ is the natural logarithm of the golden ratio $\varphi \equiv (1 + \sqrt{5})/2 \approx 1.6180$ and the subindex refers to the odd and even generations of Fibonacci numbers.
In \cite{rm23}, the study of the one-parameter Darboux deformation of (\ref{odeF}) has been motivated by the idea of introducing parametric Darboux-deformed Fibonacci numbers by taking into account that the hyperbolic function solutions are the odd and even Binet expressions of the Fibonacci numbers as previously pointed out by Faraoni and Atieh \cite{FA}. 

\medskip

The rest of the paper is structured as follows. Section II contains a brief review of the parametric supersymmetric approach. The application to equation (\ref{odeG}) is described in section III. The associated Ermakov problem and the corresponding Ermakov-Lewis invariants are discussed in section IV. Finally, some discussions and conclusions end up the paper.

\medskip

\section{Brief review of the one-parameter Darboux transformation}
\setcounter{equation}{0}

The operatorial form of (\ref{odeG}) can be written more generally as
\be\label{SchrFib}
\left(D^2+ f(t) \right) F=0~,\qquad D^2=\frac{d^2}{dt^2}~,  \qquad f(t)=\omega_0^2~,
\ee
which can be factored in the form
\be\label{SchrFib1}
(D-\Phi)(D+\Phi)F\equiv \big[D^2+(\Phi'-\Phi^2)\big]F=0~,
\ee
where $\Phi$ is the negative logarithmic derivative, $\Phi=-F'/F$, of a solution $F$ of (\ref{SchrFib}), usually known as the seed solution of the Darboux transformation. 
In the $\Phi$ variable, (\ref{SchrFib}) is turned into the Riccati equation,
\be\label{RicFib}
\Phi'-\Phi^2=f(t)~.
\ee
Way backwards, if one knows a Riccati solution $\Phi$ of (\ref{RicFib}), the solution of (\ref{SchrFib}) can be obtained from $F={\rm exp}(-\int^t \Phi)$.

\medskip

The non-parametric Darboux-transformed equation of (\ref{SchrFib1}), also known as the supersymmetric partner equation of (\ref{SchrFib1}), is obtained by reverting the factoring
\be\label{SchrFibDT}
(D+\Phi)(D-\Phi)\hat{F}\equiv \big[D^2+(-\Phi'-\Phi^2)\big]\hat{F}=\big[D^2+(f-2\Phi')\big]\hat{F}=0~.
\ee
The generic interesting fact of the reverted factorizations is that they are not unique \cite{pM,pF,RMC}.
This is because in the factorization brackets one can use the general Riccati solution in the form of the Bernoulli ansatz $\Phi_g=\Phi+1/u$, where the function $u$ satisfies the first-order differential equation
\be\label{SchrFibDTl}
-u'+2\Phi u+1=0
\ee
and not just a particular solution $\Phi$. Then, it is easy to show that
\be\label{SchrFibDTp}
\left(D+\Phi+\frac{1}{u}\right)\left(D-\Phi-\frac{1}{u}\right)\tilde{F}
= \big[D^2+(-\Phi'-\Phi^2)\big]\tilde{F}=0~.
\ee
Furthermore, the left hand side of the latter equation can be written as
\be\label{SchrFibDTp2}
\big[D^2+\left(-\Phi'_{g}-\Phi^2_{g}\right)\big]\tilde{F}=0~,
\ee
The relevant result is obtained when one reverts back the factorization brackets in the intent to return to the initial equation
\be\label{SchrFibDTp3}
\big[D^2+\left(\Phi'_{g}-\Phi^2_{g}\right)\big]{\cal F}=0~.
\ee
Substituting the Bernoulli form of $\Phi_{g}$ in (\ref{SchrFibDTp3}) leads to
\be\label{SchrFibDTp4}
\big[D^2+\left( f(t)-4\Phi u^{-1}-2u^{-2}\right)\big]{\cal F}=0~,
\ee
which is clearly different from (\ref{SchrFib1}) representing a one-parameter family of equations that have the same Darboux-transformed partner, the running parameter of the family being the integration constant that occurs in  Bernoulli's function $1/u$. The latter can be obtained by the integration of (\ref{SchrFibDTl})
\be\label{SchrFibDTp5}
\frac{1}{u}=\frac{e^{-2\int^t \Phi(t')dt'}}{\gamma +\int^t e^{-2\int^t \Phi(t')dt'}dt}=\frac{d}{dt}\ln\left(\gamma +\int^t e^{-2\int^t \Phi(t')dt'}\right)~.
\ee
Equation (\ref{SchrFibDTp4}) differs from the initial equation (\ref{SchrFib}) by
\be\label{Darbdef}
-(4\Phi/u+2/u^2)=2\frac{1}{u}\frac{d}{dt}\ln \left(\frac{1}{u}\right)~,
\ee
which is the Darboux deformation/distorsion of the nonoperatorial part $f(t)$ introduced by the parametric Darboux transformation.
Furthermore, the solutions ${\cal F}$ of the parametric Darboux-transformed equations are related to the initial (undeformed) solutions $F$ as follows
\be\label{SchrFibDTp6}
{\cal F}=e^{-\int^t \Phi_{g} ds}=e^{-\int^t \Phi ds}e^{-\int^t\frac{d}{dt}\ln\left(\gamma +\int^t e^{-2\int^t \Phi(s)ds}\right)}
=\frac{F}{\gamma +\int^t F^2(s)ds}~.
\ee

\medskip

We are now ready to apply this simple mathematical scheme to the two trigonometric cases for $f(t)=\omega_0^2$ in the next section. 

\medskip

\section{The one-parameter Darboux transformation of the classical harmonic oscillator}\label{sec-opD}
\setcounter{equation}{0}

In this section, we present the two cases that we call the odd and even one-parameter Darboux-transformed classical harmonic oscillators, corresponding to choosing the seed solution as $F\equiv F_o=\sin(\omega_0t)$ and $F\equiv F_e=\cos(\omega_0t)$, respectively.

\medskip

\subsection{The odd case.}

\medskip

\noindent In the odd case, setting $F_o=\sin(\omega_0 t)$, the Riccati solution is $\Phi_o=-\omega_0\,{\rm cot}(\omega_0 t)$ and the Darboux pair of Riccati equations has the form
\be\label{ric2}
\begin{split}
&\frac{d\Phi_o}{dt}-\Phi_o^2=\omega_0^2~, \\
 -&\frac{d\Phi_o}{dt}-\Phi_o^2=\omega_0^2[1-2{\rm cosec}^2(\omega_0t)]~,
\end{split}
\ee
where the periodic singular free term in the right hand side of the second Riccati equation is the non-parametric Darboux deformation of $\omega_0^2$.

\ms

The one-parameter Darboux partner equation (\ref{SchrFibDTp4}) takes the form of the following parametric ($\omega=\omega(t)$) oscillator equation
\be\label{peqodd}%
\begin{split}
&\frac{d^2{\cal F}_o}{dt^2}+ \omega_{o}^2(t;\gamma) {\cal F}_o=0~,\\
&\omega_{o}^2(t;\gamma)=\omega_0^2\left(1+\frac{4\sin(2\omega_0 t)}{\omega_0[2 \gamma +\tau_1(t)]}-
\frac{8\sin^4(\omega_0 t)}{\omega_0^2[2 \gamma +\tau_1(t)]^2}\right)~
\end{split}
\ee
with linearly independent solutions
\be
\begin{split} \label{3.17}
&{\cal F}_{o,1}(t;\gamma)={\cal A}_1(t;\gamma)\sin(\omega_0t)~, \quad {\cal A}_1(t;\gamma)=\frac{2}{2 \gamma +\tau_1(t)}~, \\
&{\cal F}_{o,2}(t;\gamma))={\cal A}_2(t;\gamma)\sin(\omega_0t)~, \quad {\cal A}_2(t;\gamma)=\frac{2 \omega_0 (2 \gamma +t)^2 {\rm cot} (\omega_0  t)-\tau_2(t) }{4\omega_0 \left[2 \gamma +\tau_1(t)\right]}~,  
\end{split}
\ee
where $\tau_1(t)=t[1-{\rm sinc}(2\omega_0t)]$ and $\tau_2(t)=t[1+{\rm sinc}(2\omega_0t)]$.

\ms

The first solution is obtained from (\ref{SchrFibDTp6}) and the second one from the reduction of order formula by imposing the same Wronskian, $W=-\omega_0$, as that of the undeformed pair $\{\sin(\omega_0t),\cos(\omega_0t)\}$.
They have the following symmetric and antisymmetric properties with regard to the pair of first and third quadrants
of the $(t,\gamma)$ plane
\be \label{3.18a} 
{\cal F}_{o,1}(t; \gamma)={\cal F}_{o,1}(-t; -\gamma)~, \qquad {\cal F}_{o,2}(t; \gamma)=-{\cal F}_{o,2}(-t; -\gamma)~.
\ee
\begin{figure}[h!]
\begin{center}
\includegraphics[width=14cm]  {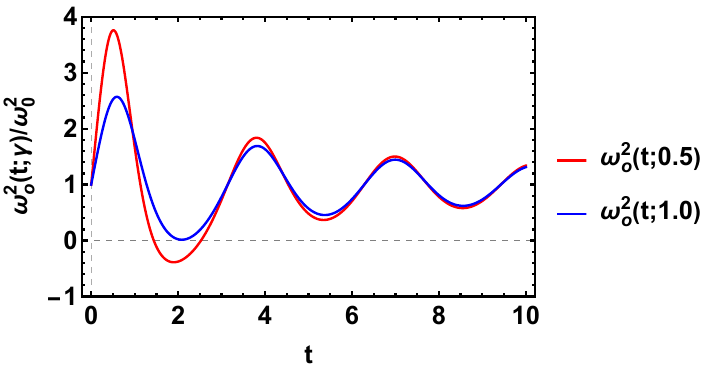}  
\includegraphics[width=14cm]  {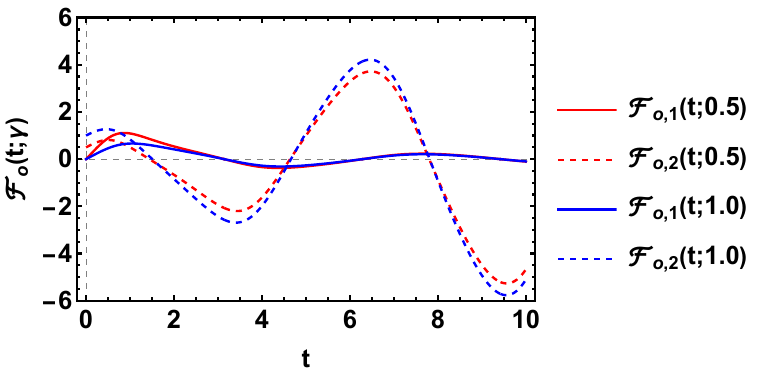}  
\caption{The square of the odd Darboux-deformed angular frequency parameter (up) and the odd deformed linearly independent solutions (down)
for $\omega_0=1$, and $\gamma=0.5$ and $1$. (In all plots of the paper, we set $\omega_0=1$).}\label{fig2}
\end{center}
\end{figure}
Plots of the angular frequency parameter of (\ref{peqodd}) and the modes ${\cal F}_{o,i}(t; \gamma)$ are given in Fig.~\ref{fig2} for the values of the parameter $\gamma$ of 0.5 and 1.

\ms

\begin{figure}[h!]
\begin{center}
\includegraphics[width=12cm] {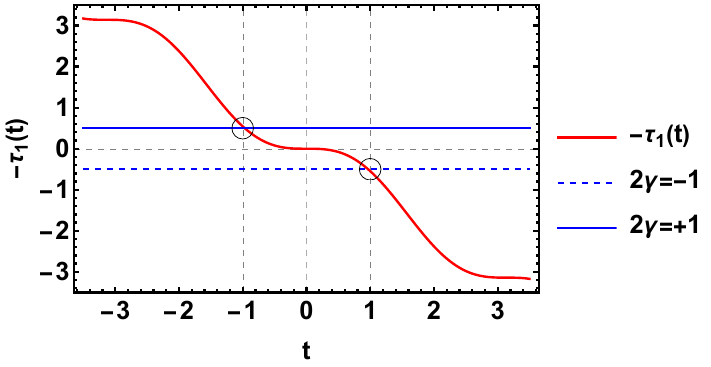} 
\includegraphics[width=12cm] {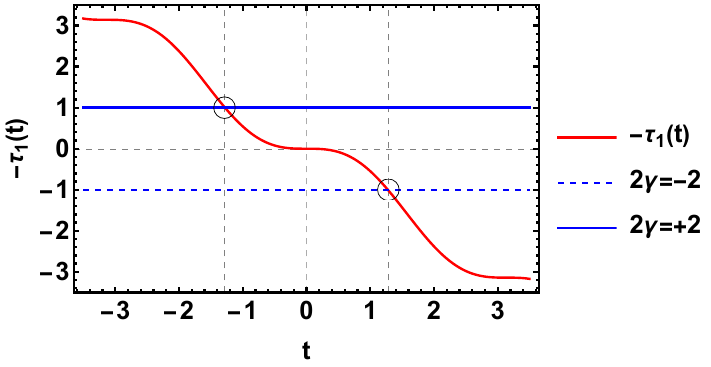} 
\caption{The coordinates on the time axis of the intersections $\gamma=-\tau_1/2$ for $\omega_0=1$, and $\gamma=\pm 0.5$ and $\pm 1$
provide the locations of the singularities for these cases.}\label{fig1}
\end{center}
\end{figure}

 The oscillator (\ref{peqodd}) is singular, and both linearly independent modes inherit the singularity at the time $t_s$ obtained as the abscissas of the roots of the transcendental equation $\gamma=-\tau_1/2$. For $\gamma>0$, the singularity is on the negative half-line (in the past), while for $\gamma<0$, it is on the positive half-line (in the future). Because of the symmetries (\ref{3.18a}), the singularities are symmetric with respect to the origin for given $|\gamma|$. In Fig.~\ref{fig1}, we present the graphical solutions obtained as the abscissas of the intersections of $\gamma$ and $-\tau_1/2$ in the cases $\gamma = \pm 05$ and $\pm 1.0$, which provide $t_s=\mp 1.27$ and $t_s=\mp 1.78$, respectively.

The square of the angular frequency (\ref{peqodd}) of this parametric oscillator starts from $\omega_0^2$ at $t=0$ with a strong oscillation which damps off rapidly to the same $\omega_0^2$ at larger times, an indicative of a transient phenomenon. As samples of this behavior, see its plots for $\gamma=$ 0.5 and 1.0 in the up side part of Fig.~\ref{fig2}. The amplitude of the first oscillation may be so strong that the oscillator can have $\omega_{o}^2(t;\gamma)<0$ in the negative sector of the initial oscillation for a lapse of time delineated by the inequality
\be \label{3.19} 
\frac{\omega_{o}^2(t;\gamma)}{\omega_{0}^2}=1 +\frac{4\sin(2\omega_0 t)}{\omega_0[2 \gamma +\tau_1(t)]} - \frac{8\sin^4(\omega_0 t)}{\omega_0^2[2 \gamma +\tau_1(t)]^2} < 0~.
\ee
For $\omega_0=1$, we have numerically determined that the parametric oscillator is completely trigonometric ($\omega_{o}^2(t;\gamma)>0$) for $\gamma > 0.97216$, while below this value it has $\omega_{o}^2(t;\gamma)<0$ for a short time interval, as seen in Fig.~\ref{fig2} for $\gamma=0.5$.
At higher frequencies, the time interval in which this oscillator is fully trigonometric extends towards the origin. For example, for $\omega_0=2$,
$\omega_{o}^2(t;\gamma)$ starts to be strictly positive at $\sim 0.48608$ and for $\omega_0=3$ at $\sim 0.32405$.

Thus, the simplest nontrivial solution of equation (\ref{peqodd}) describes relaxation oscillations from a singularity or source event in the past, while the corresponding linearly independent solution describes oscillations of higher and higher amplitudes amplifying towards a singularity in the future.
For an observer with a detector that starts measurements at the origin, only the damped solution ${\cal F}_{o,1}$ may be considered as a physical solution. However, the growing solution ${\cal F}_{o,2}$ can be used to construct the Ermakov-Lewis invariant as we will do in section \ref{sec-EPeq}.

\medskip

\subsection  {The even case.}

\medskip

\noindent In the even case, $F_e=\cos(\omega_0t)$, the Riccati solution $\Phi_e=\omega_0\tan(\omega_0t)$ leads to the Darboux pair of Riccati equations
\be\label{ric2}
\begin{split}
&\frac{d\Phi_e}{dt}-\Phi_e^2=\omega_0^2~, \\
-&\frac{d\Phi_e}{dt}-\Phi_e^2=\omega_0^2[1-2{\rm sec}^2(\omega_0 t)]~.
\end{split}
\ee
The one-parameter Darboux partner equation reads
\be\label{peqeven}
\begin{split}
&\frac{d^2{\cal F}_e}{dt^2}+ \omega_{e}^2(t;\gamma) {\cal F}_e=0~,\\
&\omega_{e}^2(t;\gamma)=\omega_0^2\left(1-\frac{4\sin(2\omega_0 t)}{\omega_0[2\gamma+\tau_2(t)]}-
\frac{8\cos^4(\omega_0 t)}{\omega_0^2[2\gamma+\tau_2(t)]^2}\right)~,
\end{split}
\ee
with the two linearly independent solutions given by
\begin{equation}\label{dse}
\begin{split}
&{\cal F}_{e,1}(t;\gamma)={\cal B}_1(t;\gamma)\cos(\omega_0t)~, \quad {\cal B}_1(t;\gamma)=\frac{2}{2 \gamma +\tau_2(t)}~, \\
&{\cal F}_{e,2}(t;\gamma)={\cal B}_2(t;\gamma)\cos(\omega_0t)~, \quad {\cal B}_2(t;\gamma)=\frac{2 \omega_0 (2 \gamma +t)^2 \tan (\omega_0  t)+\tau_1(t) }{4\omega_0 \left[2 \gamma +\tau_2(t)\right]}~,
\end{split}
\end{equation}
where again the second solution has been obtained from the reduction of order formula with the Wronskian $W=\omega_0$ corresponding to the non-deformed pair ordered as $\{\cos(\omega_0t), \sin(\omega_0t)\}$.

\begin{figure}[h!]
\begin{center}
\includegraphics[width=14cm] {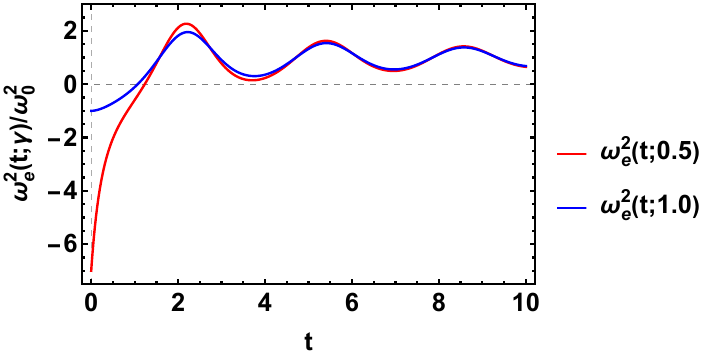} 
\includegraphics[width=14cm] {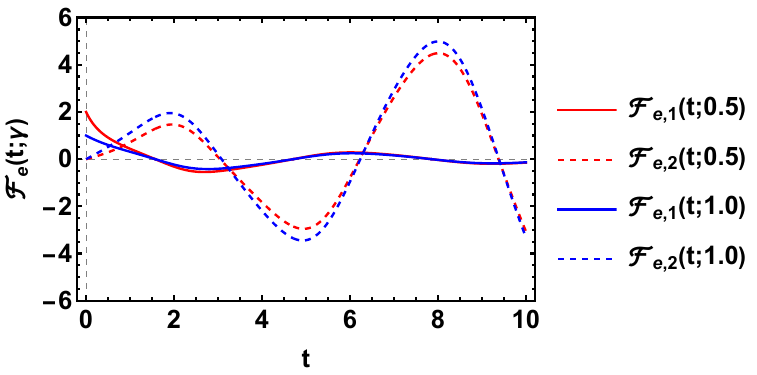} 
\caption{The same as in Fig. \ref{fig2}, but for the even case.
}\label{figfour}
\end{center}
\end{figure}

The quadrant symmetry of these modes is reversed with respect to the odd case, namely
$$
{\cal F}_{e,1}(t; \gamma)=-{\cal F}_{e,1}(-t; -\gamma)~, \qquad {\cal F}_{e,2}(t; \gamma)={\cal F}_{e,2}(-t; -\gamma)~.
$$
Plots of the angular frequency parameter in (\ref{peqeven}) and the modes ${\cal F}_{e,i}(t; \gamma)$ are given in Fig.~\ref{figfour} for the one half and unit values of the parameter $\gamma$.

\begin{figure}[h!]
\begin{center}
\includegraphics[width=12cm]  {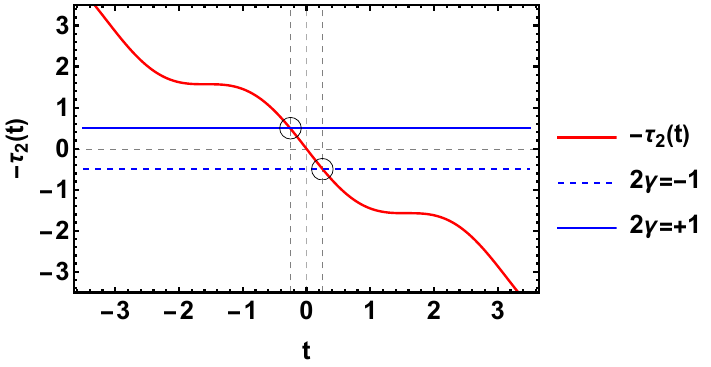}  
\includegraphics[width=12cm]  {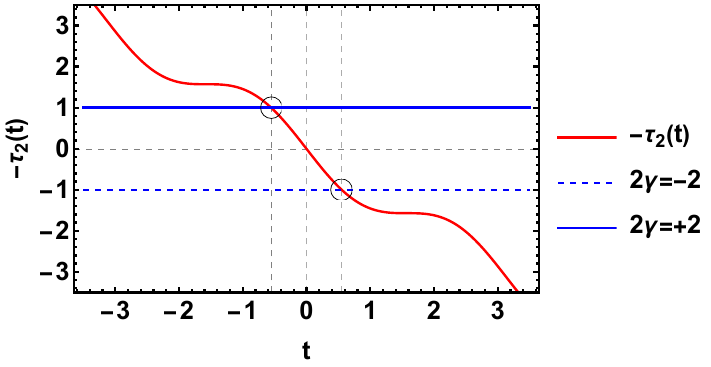}  
\caption{Same as in Fig.~\ref{fig1}, but for $\gamma=-\tau_2/2$.}\label{fig5}
\end{center}
\end{figure}

This even parametric oscillator is also singular, but now with the singularity at the time $t_s$ obtained as the root of $\gamma=-\tau_2/2$. Like in the odd case, the location of the singularities depends on the sign of $\gamma$ and have the mirror symmetry with respect to the origin at given
$|\gamma|$. In Fig.~\ref{fig5}, we present the graphical solutions obtained as the abscissas of the intersections of $\gamma$ and $-\tau_2/2$ in the
cases $\gamma = \pm 05$ and $\pm 1.0$, which provide $t_s=\mp 0.55$ and $t_s=\mp 2.48$, respectively.

Examination of the square of the angular frequency $\omega_e^2$ shows that at $t=0$ it is $\omega_e^2(0;\gamma)=2/\gamma^2-\omega_0^2$. Thus for $\gamma<\sqrt{2}/\omega_0$ this parametric oscillator is of trigonometric type at least in some small time interval of the origin. However, the situation
of the amplitude of the first oscillation on the positive time line is more involved than in the odd case and results from the analysis of the roots of
the transcendental equation
\be \label{treq}
\frac{\omega_{e}^2(t;\gamma)}{\omega_{0}^2}=1-\frac{4\sin(2\omega_0 t)}{\omega_0[2\gamma+\tau_2(t)]}-\frac{8\cos^4(\omega_0 t)}{\omega_0^2[2\gamma+\tau_2(t)]^2} > 0~.
\ee
Based on (\ref{treq}), in the case $\omega_0=1$, we determined that for $\gamma > 1.75750$, the even oscillator has always a positive angular frequency parameter. However, as shown in the up-sided plots of Fig.~\ref{figfour}, the frequency parameter $\omega_{e}^2(t;\gamma)$ can be negative in some time intervals, and
these oscillations of the angular parameter generate the oscillatory solutions (\ref{dse}) as presented in the down-sided plots of Fig.~\ref{figfour}.
We have determined numerically the following pattern for values of $\gamma$ below 1.7575. For $\gamma\in (0,0.18677)$, $\omega_{e}^2(t;\gamma)$ has a sequence of $(-,+,-)$ values before turning completely positive. For $\gamma\in (0.18677, 1.41421)$, there is only an initial negative region after which
$\omega_{e}^2(t;\gamma)>0$ always. When $\gamma\in (1.41421, 175750)$, there exist a sequence of $(+,-)$ values of $\omega_{e}^2(t;\gamma)$ before it turns  fully positive.

For $\omega_0=2$, the same $\gamma$ intervals are reduced as follows: $(0, 0.0934)$, $(0.0934,0.70710)$, $(0.70710,0.87879)$. Beyond $\gamma=0.87879$, $\omega_{e}^2(t;\gamma)$ is always positive.

For $\omega_0=3$, the intervals are further reduced to: $(0, 0.0625)$, $(0.0625, 0.4714)$, $0.4714,0.5875)$,
and beyond $\gamma=0.5875$, the even angular frequency parameter is strictly positive.

By the same token as in the odd case, the damped oscillatory solutions may be considered as the physical ones. However, the growing oscillatory solutions are still useful in the construction of the Ermakov-Lewis invariants. 

\medskip

\section{The associated Ermakov-Pinney equations and the Ermakov-Lewis invariants}\label{sec-EPeq}
\setcounter{equation}{0}

The Ermakov-Pinney equations are nonlinear extensions of the linear second-order differential equations whose solutions together with the corresponding linear solutions are used to build the Ermakov-Lewis dynamical invariants. We first consider the original harmonic oscillator equation in this linear-nonlinear context and then present the odd and even parametric Darboux-transformed equations in the same framework.

\subsection{The harmonic oscillator case.}

We consider (\ref{odeG}) as a parametric oscillator equation in the particular cases of constant coefficients, which has been studied in more detail in \cite{maro}.
It is a well-known result that one can use two given linear independent solutions, $u_1$, and $u_2$, of the parametric oscillator equation,
\begin{equation}\label{a9}
u'' +\omega^2(t)u=0~,
\end{equation}
to build a particular solution of the corresponding nonlinear Ermakov-Pinney equation
\begin{equation}\label{a10}
v''+\omega^2(t)v+kv^{-3}=0~,
\end{equation}
where $k\neq 0$ is an arbitrary real constant, by means of Pinney's formula \cite{P}
\begin{equation}\label{a11}
v(t)=\sqrt{u_1^2-\frac{ku_2^2}{W^2}}~.
\end{equation}
where $W$ is the Wronskian of the two linearly independent solutions $u_1$ and $u_2$.

\medskip

For the case $\omega^2(t)=\omega_0^2$, a constant, the Ermakov-Pinney equation has the autonomous form \cite{maro}
\begin{equation}\label{a13}
v''+\omega_0^2v+kv^{-3}=0~.
\end{equation}
From (\ref{a11}), by taking $u_1=\sin(\omega_0t), ~u_2=\cos(\omega_0t)$, and $W=-\omega_0$, the particular solution of \eqref{a13} is
\be\label{a14vo}
v_o(t)=\sqrt{\sin ^2(\omega_0 t)-\kappa \cos ^2(\omega_0 t )}=\sqrt{1-\kappa\cot ^2(\omega_0 t)}\,\sin(\omega_0 t)~,
\ee
where $\kappa=k/\omega_0^2$.

\begin{figure}[h!]
\begin{center}
\includegraphics[width=14cm]  {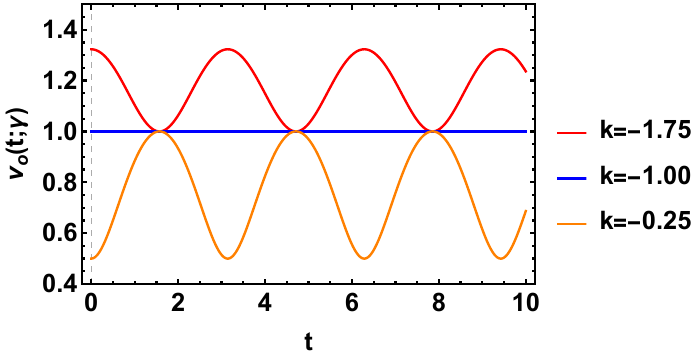} 
\includegraphics[width=14cm]   {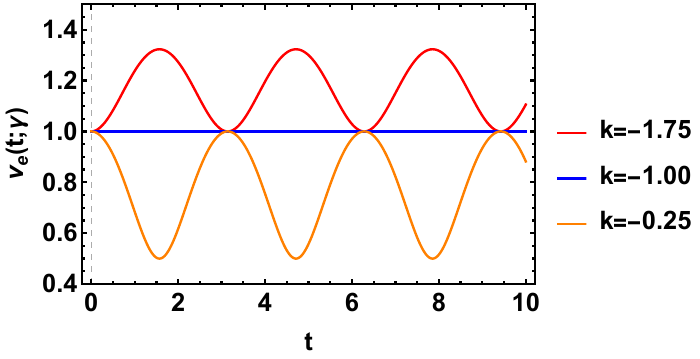} 
\caption{The non-defomed Ermakov-Pinney solutions (\ref{a14vo}) and (\ref{a14ve}) for $\kappa=-1.75, -1, -0.25$.}
\label{fvsmall}
\end{center}
\end{figure}

On the other hand, if one takes $u_1=\cos(\omega_0t), ~u_2=\sin(\omega_0t)$, and $W=\omega_0$, the particular solution of \eqref{a13} is
\be\label{a14ve}
v_e(t)=\sqrt{\cos ^2(\omega_0 t)-\frac{k \sin ^2(\omega_0 t )}{\omega_0^2}}=\sqrt{1-\kappa\tan ^2(\omega_0 t)}\,\cos(\omega_0 t)~.
\ee
Notice that these EP solutions are written in the  amplitude-modulated form of the linear solutions. Plots for three values of the nonlinear parameter $k$ are given in Fig.~\ref{fvsmall}.

Both (\ref{a14vo}) and (\ref{a14ve}) can be used to calculate the Ermakov-Lewis invariant defined as
\be\label{einv}
{\cal I}_{v_o}=\frac{1}{2}\bigg[-k\Big(\frac{u}{v_o}\Big)^2+(u'v_o-uv_o')^2\bigg]~, \qquad {\cal I}_{v_e}=\frac{1}{2}\bigg[-k\Big(\frac{u}{v_e}\Big)^2+(u'v_e-uv_e')^2\bigg]~.
\ee
Using for $u$ the linear superposition $u=M u_1+N u_2$ in each of the latter expressions,
the Ermakov-Lewis invariants become \cite{mr2014}
\be\label{einv}
{\cal I}_{v_o}={\cal I}_{v_e}=\frac{1}{2}\left(\omega_0^2N^2-kM^2\right)=const~.
\ee

\subsection{The one-parameter Darboux-deformed cases.}

For the one-parameter Darboux-deformed counterparts, the Ermakov-Pinney equations are

\begin{equation}\label{vcal}
{\cal V}_{o}''+\omega_{\gamma,o}^2{\cal V}_{o}+k{\cal V}_{o}^{-3}=0~, \quad
{\cal V}_{e}''+\omega_{\gamma,e}^2{\cal V}_{e}+k{\cal V}_{e}^{-3}=0~.
\end{equation}

In the odd case, we take  $u_1={\cal F}_{o,1}(\omega_0t;\gamma)$ and $u_2={\cal F}_{o,2}(\omega_0t;\gamma)$ from \eqref{3.17}
and using \eqref{a11}, we have
\begin{equation}\label{vodd1}
\mathcal{V}_o(t;\gamma)=\pm\sqrt{1-\kappa_o(t;\gamma)\cot ^2(\omega_0 t)}\,{\cal F}_{o,1}(t;\gamma)~,
\end{equation}
where
$$
\kappa_o(t;\gamma)=\frac{kV_o(t;\gamma)}{\omega_0^2}~, \qquad V_o(t;\gamma)=\bigg[\frac{\tau_2(t)\tan(\omega_0t)-2\omega_0(2\gamma+t)^2}{8\omega_0}\bigg]^2~.
$$
Let us notice that $\mathcal{V}_o$ is not singular for $t\geq 0$ if $\gamma>0$.

\medskip

For the even case, taking  $u_1={\cal F}_{e,1}(t;\gamma)$ and $u_2={\cal F}_{e,2}(t;\gamma)$ from \eqref{dse},
Pinney's formula gives
\begin{equation}\label{veven1}
\mathcal{V}_e(t;\gamma)=
\pm\sqrt{1-\kappa_e(t;\gamma)\tan ^2(\omega_0 t)}\,{\cal F}_{e,1}(t;\gamma)~,
\end{equation}
where
$$
\kappa_e(t;\gamma)=\frac{kV_e(t;\gamma)}{\omega_0^2}~, \qquad V_e(t;\gamma)=\bigg[\frac{\tau_1(t)\cot(\omega_0t)+2\omega_0(2\gamma+t)^2}{8\omega_0}\bigg]^2~.
$$
Likewise the odd case, $\mathcal{V}_e(t;\gamma)$ is not singular for $t\geq 0$ if $\gamma>0$.

\begin{figure}[h!]
\begin{center}
\includegraphics[width=14cm] {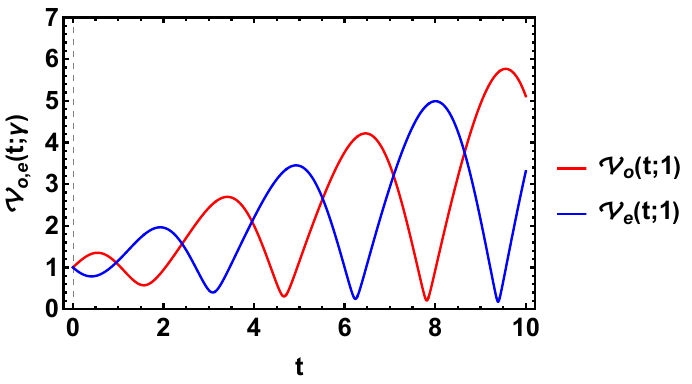} 
\includegraphics[width=14cm] {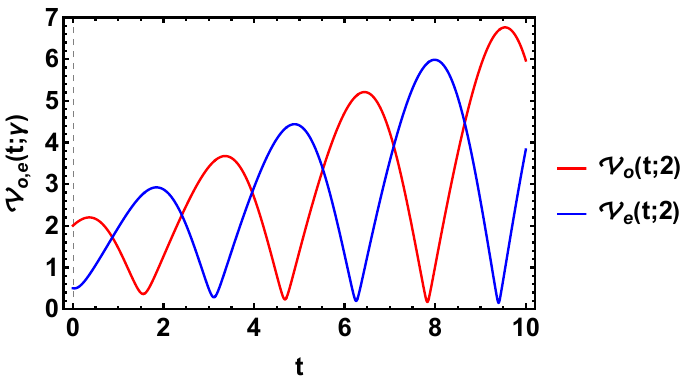} 
\caption{The Darboux-deformed odd and even Ermakov-Pinney solutions with $k=-1$ for $\gamma=1$ (up) and $\gamma=2$ (down).}\label{figdv}
\end{center}
\end{figure}

Plots of the parametric Ermakov-Pinney solutions ${\cal V}_o$ and ${\cal V}_e$ for $\gamma=1$ and $\gamma=2$ are displayed in Fig.~\ref{figdv}.

\medskip

The Ermakov-Lewis invariant is
\be\label{einv1}
{\cal I}_{{\cal V}_{o}}=\frac{1}{2}\bigg[-k\Big(\frac{u}{{\cal V}_o}\Big)^2+(u'{\cal V}_o-u{\cal V}_o')^2\bigg]~, \quad
{\cal I}_{{\cal V}_{e}}=\frac{1}{2}\bigg[-k\Big(\frac{u}{{\cal V}_e}\Big)^2+(u'{\cal V}_e-u{\cal V}_e')^2\bigg]~.
\ee

\begin{figure}[h!]
\begin{center}
\includegraphics[width=14cm]{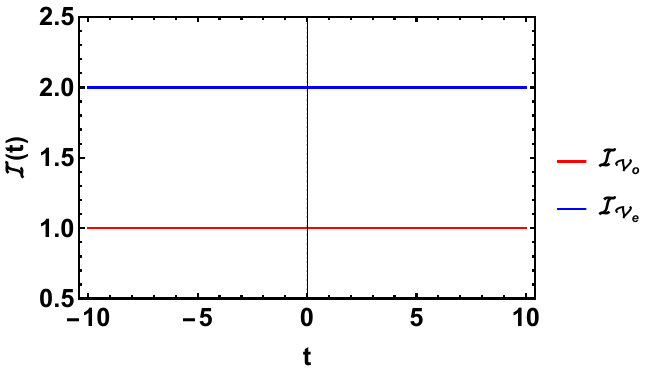} 
\caption{Ermakov-Lewis invariants: the odd case (red) is for $\tilde{M}=\tilde{N}=1$ and $k=-1$ and the even case (blue) is for $\tilde{M}=\sqrt{15}, \tilde{N}=\frac{1}{2}$ and $k=-\frac{1}{4}$.}\label{f5}
\end{center}
\end{figure}

Similarly to the undeformed case, we let now $u=\tilde{M} {\cal F}_{o,1}+\tilde{N} {\cal F}_{o,2}$ with ${\cal V}_o$ from \eqref{vodd1}, and $u=\tilde{M} {\cal F}_{e,1}+\tilde{N} {\cal F}_{e,2}$ with ${\cal V}_e$ from \eqref{veven1}. After straightforward calculations, the invariants become
\be\label{einv2}
\mathcal{I}_{{\cal V}_{o}}=\frac{1}{2}\left(\omega_0^2\tilde{N}^2-k\tilde{M}^2\right)=const~,~~\mathcal{I}_{{\cal V}_{e}}=\frac{1}{2}\left(\omega_0^2\tilde{N}^2-k\tilde{M}^2\right)=const~.
\ee
We have checked the constant values of the invariants given in (\ref{einv2}) by the plots of its expressions in (\ref{einv1}) as displayed in Fig.~\ref{f5}. The expressions in Eq.~(\ref{einv2}) show, as expected, that the Ermakov-Lewis invariants do not depend on the Darboux deformation parameter.

\medskip

\begin{figure}[h!]
\begin{center}
\includegraphics[width=14cm]{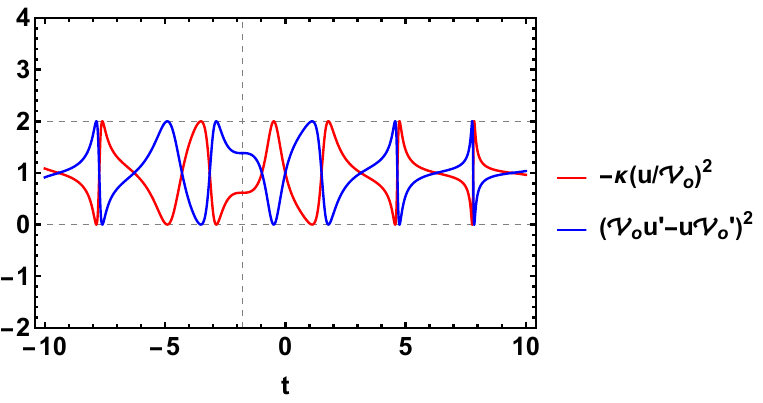}
\includegraphics[width=14cm]{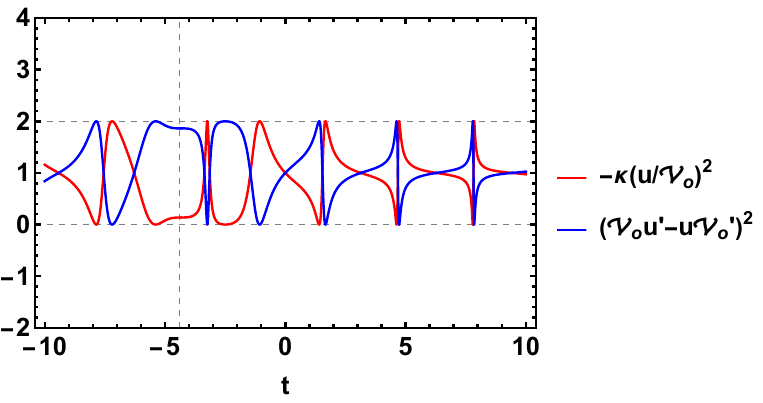}
\caption{The oscillations of the two terms in the Ermakov-Lewis invariants in the first case from the previous figure.
The vertical dashed lines indicate the time location of the parameter-induced singularities in the deformed functions and the corresponding Pinney functions at $t=-1.78$ and $t=-4.34$ for $\gamma=1$ and $\gamma=2$, respectively.
}\label{figosc1}
\end{center}
\end{figure}

\begin{figure}[h!]
\begin{center}
\includegraphics[width=14cm] {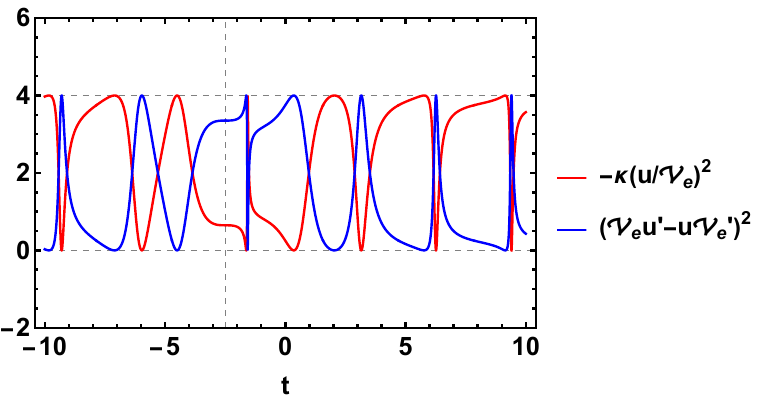} 
\includegraphics[width=14cm]  {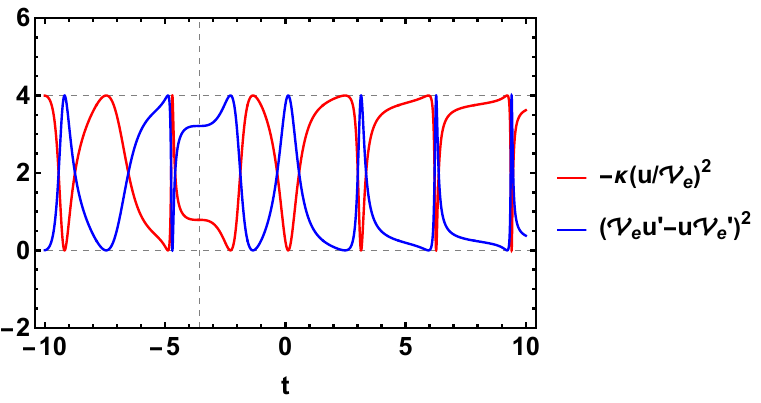} 
\caption{The oscillations of the two terms in the Ermakov-Lewis invariants in the even case from Fig.~\ref{f5}.
The vertical dashed lines indicate the time location of the parameter-induced singularities in the deformed functions and the corresponding Pinney functions at $t=-2.484$ and $t=-3.602$ for $\gamma=1$ and $\gamma=2$, respectively.
}\label{figosc2}
\end{center}
\end{figure}

Another interesting feature of the Ermakov-Lewis invariant for the Darboux-deformed parametric oscillators is that it is free of both trigonometric
and parameter singularities. This is due to the special form of the deformed solutions and the corresponding Pinney solutions which provide opposite oscillations around the constant value of the invariant, see Figs.~\ref{figosc1} and \ref{figosc2}.

\medskip

\begin{figure}[h!]
\begin{center}
\includegraphics[width=14cm] {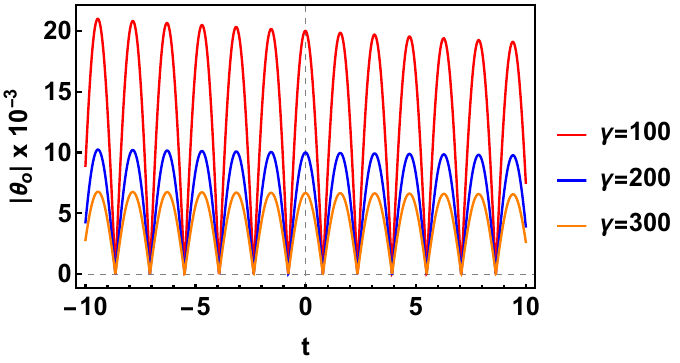} 
\includegraphics[width=14cm] {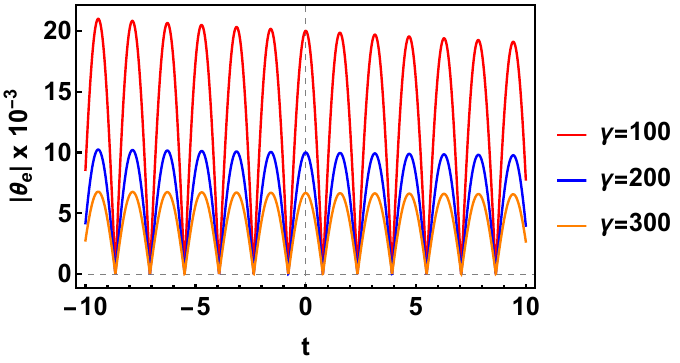}   
\caption{The adiabaticity parameter $|\theta|$ for the one-parameter Darboux-deformed modes ${\cal F}_{o1}$ and ${\cal F}_{e1}$ for various values of $\gamma$.}\label{figtheta}
\end{center}
\end{figure}

To this end, following \cite{ray1979}, we discuss the adiabatic regime of these singular parametric oscillators. This dynamical regime, defined by the parameter $|\theta|=|d{\omega/dt}|/\omega^2(t)\ll 1$, is an important technological characteristic of parametric oscillators and is controlled by its
own quasi-invariant known as the adiabatic invariant
\begin{equation}\label{adinv}
I_{{\rm ad}}=\frac{E(t)}{\omega(t)}~,
\end{equation}
where the energy of the oscillator is given by
\begin{equation}\label{oscenerg}
E(t)=\frac{1}{2}\big[ (d{\cal F}/dt)^2+\omega^2(t){\cal F}^2\big]~.
\end{equation}
Lewis discovered that the adiabatic `invariant' is the leading term in the adiabatic series expansion of the Ermakov-Lewis invariant \cite{lewis1968}.
However, an adiabatic regime for the parametric Darboux-deformed modes can exist only
for the modes ${\cal F}_{o1}$ and ${\cal F}_{e1}$ since their energy is decreasing in time, but the adiabaticity condition on the parameter $\theta$ is better satisfied only at high values of the parameter $\gamma$, see Fig.~\ref{figtheta}, where the plots show that $|\theta|< 2$\% for $\gamma>$ 100 in both cases.

\medskip

\section{Discussion and Conclusions}
\setcounter{equation}{0}
\label{sec:5}

In this paper, we have introduced and discussed in some detail an interesting class of singular parametric oscillators obtained through the one-parameter Darboux deformation/transformation of the classical harmonic oscillator.
In an innovative way, the oscillatory singular solutions have been written as amplitude-modulated solutions of the sine and cosine functions that have been used as seed solutions.
Moreover, for the corresponding Ermakov-Pinney equations, the solutions have been also written in the amplitude-modulated form making easier the calculation of the Ermakov-Lewis invariants. The latter invariants have been shown to depend on the superposition constants of the linearly independent harmonic solutions, but not on the deformation parameter. Significant insight has been accomplished by showing that the Ermakov-Lewis invariants remain true dynamical invariants for these Darboux-deformed parametric oscillators despite their parametric and trigonometric singularities. This is due to the opposite functional behaviours of the quotient term and the Wronskian-like term in the formula of the Ermakov-Lewis invariant that lead to cancelations of the singularities. An adiabatic regime may be considered, but only for the decaying modes at high values of the parameter $\gamma$.

The more general case of this kind of parametric oscillators possessing constant friction or gain can be easily analysed by adding a constant shift to the Riccati solution, a procedure that we described in \cite{rk2011}, or by direct factorization as in \cite{rr98}, which is similar to the nonlinear factorizations of Rosu and Cornejo-P\'erez \cite{rcp1,rcp2}, which can be also adapted to differential equations with nonautonomous friction-gain terms.

One obvious application of the results of this paper, including those related to the Ermakov-Lewis approach, is in cosmology, where the Riccati/supersymmetry approach in the conformal time variable has been long considered in the FRW barotropic models \cite{r2000,r98,r02,rk2013} and isotropic scalar field cosmologies \cite{hlm2014}. The even case considered in the present paper corresponds to the evolution of a closed vacuum-filled FRW universe in conformal time. Regarding the usage of the Ermakov-Lewis invariants in this area, we mention the comment paper of Ray \cite{ray1979}, who related the cosmological particle production in the adiabatic regime to the Ermakov-Lewis invariant of parametric oscillator modes with the angular frequency depending on the time-dependent metric of cosmological models with an initial singularity. We also recall here the recent surge of interest in this approach in a number of stimulating, cosmology-sided papers \cite{bkpt24,crs24,barb06}, but with some results that can be adapted to the quantum-to-classical transition at microscopic scales.

 \bigskip
\newpage

\noindent {\bf Credit author statement}

\medskip

H.C. Rosu: Writing - original draft, Supervision, Formal analysis.

J. de la Cruz: Calculation, Investigation.

\bigskip

\noindent {\bf Declaration of Competing Interest}

\medskip

The authors declare that they have no known competing financial interests or personal relationships that could have appeared
to influence the work reported in this paper.

\bigskip

\noindent {\bf Data availability}

\medskip

No data was used for the research described in the paper.

\bigskip
\bigskip

\begin{center} {\bf Acknowledgments} \end{center}

The authors thank the referee for inspiring remarks that help them to greatly improve the quality of this research. The second author acknowledges the financial support of CONAHCyT-Mexico through a doctoral fellowship at the early stage of this work.



\begin{thebibliography}{99}

\bibitem{pM} B. Mielnik, Factorization method and new potentials with the oscillator spectrum, J. Math. Phys. 25 (1984) 3387-3389.
 https://doi.org/10.1063/1.526108

\bibitem{pF} D.J. Fern\'andez C., New hydrogen-like potentials, Lett. Math. Phys. 8 (1984) 337-343.
 https://doi.org/10.1007/BF00400506

\bibitem{psp} J. Pappademos, U. Sukhatme, A. Pagnamenta, Bound states in the continuum from supersymmetric quantum mechanics,
Phys. Rev. A 48 (1993) 3525-3531.
 https://doi.org/10.1103/PhysRevA.48.3525 

 \bibitem{cks} F. Cooper, A. Khare, U. Sukhatme, Supersymmetry and quantum mechanics,
 Phys. Rep. 251 (5-6) (1995) 267-385.
 https://doi.org/10.1016/0370-1573(94)00080-M

\bibitem{rosu96-1} H.C. Rosu, Darboux-Witten techniques for the Demkov-Ostrovsky problem, Phys. Rev. A 54 (1996) 2571-2576.
 https://doi.org/10.1103/PhysRevA.54.2571

\bibitem{rosu96-2} H.C. Rosu, J. Socorro, One-parameter family of closed radiation-filled FRW ``quantum" universes, Phys. Lett. A 223 (1996) 28-30.
 https://doi.org/10.1016/S0375-9601(96)00709-8

\bibitem{rosu97} H.C. Rosu, Supersymmetric Fokker-Planck strict isospectrality, Phys. Rev. E 56 (1997) 2269-2271.
 https://doi.org/10.1103/PhysRevE.56.2269

\bibitem{r2000} H.C. Rosu, Darboux class of cosmological fluids with time-dependent adiabatic indices,
Mod. Phys. Lett. A 15 (2000) 979-989.
 https://doi.org/10.1142/S0217732300000980

\bibitem{RMC} H.C. Rosu, S.C. Mancas, P. Chen,
One-parameter families of supersymmetric isospectral potentials
from Riccati solutions in function composition form, Ann. Phys. 343 (2014) 87-102.
 https://doi.org/10.1016/j.aop.2014.01.012

 \bibitem{KKYK} R. Kumar, R. Kumar Yadav,A. Khare,
 Rationally extended harmonic oscillator potential, isospetral family and the uncertainty relations,
  Ann. Phys. 463 (2024) 169623.
 https://doi.org/10.1016/j.aop.2024.169623


\bibitem{rr98} H.C. Rosu, M.A. Reyes, Riccati parameter modes from Newtonian free damping motion by supersymmetry,
Phys. Rev. E 57 (1998) 4850-4852.
 https://doi.org/10.1103/PhysRevE.57.4850

\bibitem{re01} H.C. Rosu, P.B. Espinoza, Ermakov-Lewis angles for one-parameter supersymmetric families of Newtonian free damping modes,
Phys. Rev. E 63 (2001) 037603.
 https://doi.org/10.1103/PhysRevE.63.037603

 \bibitem{td1} S.A. Hojman, H.M. Moya-Cessa, F. Soto-Eguibar, F.A. Asenjo, Time-dependent harmonic oscillators and SUSY in time domain,
 Phys. Scr. 96 (2021) 125218.  https://doi.org/ 10.1088/1402-4896/ac267d

 \bibitem{td2} K. Zelaya, O. Rosas-Ortiz, Exact solutions for time-dependent non-Hermitian oscillators: Classical and quantum pictures,
 Quantum. Rep. 3 (2021) 458-472.  https://doi.org/10.3390/quantum3030030

 \bibitem{td3} C. Garc\'{\i}a-Meca, A. Macho Ortiz, R. Llorente S\'aez, Supersymmetry in the time domain and its application in optics,
 Nat. Commun. 11 (2020) 813. https://doi.org/10.1038/s41467-020-14634-0

 \bibitem{bagchibook} B.K. Bagchi, {\em Supersymmetry in quantum and classical mechanics}, Chapmann and Hall/CRC, Boca Raton, 2001.

 \bibitem{bissoni86} S.N. Biswas, S.K. Soni, Supersymmetric classical mechanics,  Pramana-J. Phys. 27 (1986) 117-127.
 https://doi.org/10.1007/BF02846333

 \bibitem{ioffe06} M.V. Ioffe, J. Mateos Guilarte, P.A. Valinevich, Two-dimensional supersymmetry: From SUSY quantum mechanics to integrable classical
 models, Ann. Phys. 321 (2006) 2552-2565. https://doi.org/10.1016/j.aop.2006.02.011

\bibitem{rm23} H.C. Rosu, S.C. Mancas, One-parameter Darboux-deformed Fibonacci numbers,
Mod. Phys. Lett. A 38 (2023) 2350022. https://doi.org/10.1142/S0217732323500220


\bibitem{FA} V. Faraoni, F. Atieh,
Generalized Fibonacci numbers, cosmological analogies, and an invariant,
Symmetry 13 (2021) 200. https://doi.org/10.3390/sym13020200


\bibitem{P} E. Pinney, The nonlinear differential equation $y'' + p(x)y + cy^{-3} = 0$, Proc. Am. Math. Soc. 1 (1950) 681-681.
 https://doi.org/10.1090/S0002-9939-1950-0037979-4

\bibitem{maro} S.C. Mancas, H.C. Rosu, Integrable differential equations with Ermakov nonlinearities and Chiellini damping, Appl. Math. Comp. 259 (2015) 1-11.  https://doi.org/10.1016/j.amc.2015.02.037

\bibitem{mr2014} S.C. Mancas, H.C. Rosu, Ermakov-Lewis invariants and Reid systems, Phys. Lett. A 378 (2014) 2113-2117.
 https://doi.org/10.1016/j.physleta.2014.05.008

 \bibitem{rk2011} H.C. Rosu, K.V. Khmelnytskaya, Shifted Riccati procedure: Application to conformal barotropic FRW cosmologies,
 SIGMA 7 (2011) 013.  https://doi.org/10.3842/SIGMA.2011.013

 \bibitem{rcp1}
H.C. Rosu, O. Cornejo-P\'erez,
Supersymmetric pairing of kinks for polynomial nonlinearities, 
Phys. Rev. E 71 (2005) 046607. 
https://doi.org/10.1103/PhysRevE.71.046607

\bibitem{rcp2}
O. Cornejo-P\'erez, H.C. Rosu,
Nonlinear second order ODE's: Factorizations and particular solutions, 
Prog. Theor. Phys. 114 (2005) 533-538.
https://doi.org/10.1143/PTP.114.533

\bibitem{ray1979} J.R. Ray, Cosmological particle production, Phys. Rev. D 20 (1979) 2632-2633.
https://doi.org/10.1103/PhysRevD.20.2632

\bibitem{lewis1968} H.R. Lewis, Jr., Class of exact invariants for classical and quantum time-dependent harmonic oscillators,
J. Math. Phys. 9 (1968) 1976-1986. https://doi.org/10.1063/1.1664532

\bibitem{r98} H.C. Rosu, Darboux parameter for empty FRW quantum universes and quantum cosmological singularities,
Mod. Phys. Lett. A 13 (1998) 227-230. https://doi.org/10.1142/S0217732398000280

\bibitem{r02} H.C. Rosu, KdV adiabatic index solitons in barotropic open FRW cosmologies, Mod. Phys. Lett. A 17 (2002) 667-670.
https://doi.org/10.1142/S0217732302006898

\bibitem{rk2013} H.C. Rosu, K.V. Khmelnytskaya, Inhomogeneous barotropic FRW cosmologies in conformal time,
Mod. Phys. Lett. A 28 (2013) 1340017. https://doi.org/10.1142/S0217732313400178

\bibitem{hlm2014} T. Harko, F.S.N. Lobo, A Riccati equation based approach to isotropic scalar field cosmologies,
Int. J. Mod. Phys. D 23 (2014) 1450063.
https://doi.org/10.1142/S0218271814500631

\bibitem{bkpt24} K. Boutivas, D. Katsinis, G. Pastras, N. Tetradis, Entanglement in cosmology, JCAP04 (2024) 017.
https://doi.org/10.1088/1475-7516/2024/04/017

\bibitem{crs24} S.M. Chandran, K. Rajeev, S. Shankaranarayanan, Real-space quantum-to-classical transition of time-dependent background fluctuations,
Phys. Rev. D 109 (2024) 023503.
https://doi.org/10.1103/PhysRevD.109.023503

\bibitem{barb06} J.F. Barbero G., D. G\'omez Vergel, E.J.S. Villase\~nor, Evolution operators for linearly polaarized two-Killing cosmological models,
Phys. Rev. D 74 (2006) 024003.
https://doi.org/10.1103/PhysRevD.74.024003



\end{thebibliography}
\end{document}